\documentclass[11pt]{article}
\usepackage{moriond_my,epsfig}

\def\be{\begin{equation}}
\def\ee{\end{equation}}
\def\bea{\begin{eqnarray}}
\def\eea{\end{eqnarray}}

\def\DAF{DA$\Phi$NE }
\def\Repsp{Re($\epsilon'/\epsilon $)}

\def\p00{$\pi^{0}\pi^{0}$}
\def\KL{K$^{0}_{L}$}
\def\KS{K$^{0}_{S}$}
\def\KP{K$^{+}$}
\def\KM{K$^{-}$}

\def\K3{$K_{l3}$}
\def\K4{$K_{l4}$}
\def\lum{cm$^{-2}$s$^{-1}$}
\def\K00{K$^{0}_{S} \rightarrow \pi^{0}\pi^{0}$}
\def\Kpm{K$^{0}_{S} \rightarrow \pi^{+}\pi^{-}$}
\def\Kpmg{K$^{0}_{S} \rightarrow \pi^+\pi^-(\gamma)$}

\def\L3p{K$^{0}_{L} \rightarrow \pi^{+}\pi^{-}\pi^{0}$}
\def\Len{K$^{0}_{L} \rightarrow \pi^{\pm}e^{\mp}\nu$}
\def\Sep{K$^{0}_{S} \rightarrow \pi^{\pm}e^{\mp}\nu$}

\def\Ps0{P$_{S0}$}
\def\Pl1{P$_{L1}$}

\begin{document}
\vspace*{4cm}
\title{KLOE recents results}

\author{KLOE collaboration\footnote{
%M.~Adinolfi,
A.~Aloisio,
F.~Ambrosino,
%A.~Andryakov,
A.~Antonelli,
M.~Antonelli,
C.~Bacci,
%G.~Barbiellini,
%F.~Bellini,
G.~Bencivenni,
S.~Bertolucci,
C.~Bini,
C.~Bloise,
V.~Bocci
% ...
%P.~P.~Zhao,
%Y.~Zhou.

}}
\address{presented by V.Patera, INFN, Laboratori Nazionali di Frascati\\
via E. Fermi 40, Frascati, Italy\\}

\maketitle\abstracts{
Since april 1999, the KLOE
experiment at \DAF\  has collected about 200 pb$^{-1}$ of data,
produced in $e^{+}e^{-}$ collision at the c.m. energy of 1020
MeV, the mass of the $\phi$ meson. These data have been used
for detailed studies on the $\phi$ radiative decays, as well as on
rare \KS\ decays. The first results, based on the $\sim$ 20
pb$^{-1}$ collected in year 2000 are presented here.}

\section{Introduction}

The KLOE\cite{kp} detector at \DAF \cite{dafne}, the Frascati $\phi$-factory, 
has begun physics data taking in april 1999. The machine operates at the
peak of the $\phi$(1020) meson production cross section from
e$^{+}$-e$^{-}$ collisions with a yield of $\simeq 3 \times 10^6 \phi$
produced for every delivered pb$^{-1}$. 
With a peak design luminosity of 5$\times 10^{32}$
\lum\, \DAF could deliver as many as $40 pb^{-1}/day$, being
therefore an exceptional source of almost monochromatic \KS\ and
\KL\ pairs, \KP\ and \KM\ pairs, and all other $\phi$ decay products.

%the main goal of KLOE is the measurement of the real
%part of the direct cp violation parameter \epsp, task 
% that needs full machine luminosity for $\sim$ one year in order to reach the
%desired statistic accuracy.
%besides cp violation, however, KLOE is well suited for the study
%of many other physics topics, both in kaon and non-kaon physics, that do not 
%require the same huge statistics needed by the\epsp\ measurement.
At present \DAF\  reaches a peak luminosity of
$\simeq 5 \times 10^{31}$ \lum\ and delivers routinely $\simeq 2
pb^{-1}/day$. This is already the highest
luminosity ever reached at these energies and is the
result of a constant improvement of performances along time.
In the present paper the first KLOE results on \KS\ decays and radiative 
$\phi$ decays based on the year 2000 data are presented. 

\section{The KLOE detector}

The KLOE detector\cite{kp,tp} 
%(see figure \ref{KLOEdet}) 
consists of a large tracking chamber, and a hermetic electromagnetic
calorimeter. A large magnet, which consists of a superconducting
coil and an iron yoke surrounding the whole detector, provides an
axial magnetic field of 0.52 T.

The tracking chamber \cite{dpap} (dc) is a cylindrical, 2 m radius,
 3.3 m long drift chamber.
The total number of wires is 52140, out of which 12582 are the
sense wires. It operates with a low-Z, He gas mixture, to minimize
multiple scattering of charged particles and regeneration of \KL\. 
The 58 concentric layers of wires are strung in an all-stereo
geometry, with constant inward radial displacement at the chamber
center. A spatial resolution better than $200 \mu m$ is obtained.
The momentum resolution  for 510 MeV/c electrons and positrons is
1.3 MeV/c, in the angular range 130$^{\circ} > \theta >$
50$^{\circ}$.

The electromagnetic calorimeter\cite{tp,epap} (EmC) is a lead-scintillating
fibers sampling
calorimeter, divided into a barrel section and two endcaps. The modules of
 both sections
are read-out at the two ends by a total of 4880 photomultipliers.
In order to minimize dead zones in the overlap region between
barrel and endcaps, the modules of the latter are bent outwards
with respect to the decay region.

The calorimeter was designed to detect with very high efficiency
photons with energy as low as 20 MeV and to accurately measure
their energy and time of flight. Calibrations of energy and time
scales are performed simultaneously using appropriate data
sub-samples during data collection. An energy resolution of
5.7$\%$/$\sqrt{E(GeV)}$ is achieved throughout the whole
calorimeter, together with a linearity in energy response better
than 1$\%$ above 80 MeV and better than 4$\%$ between 20 to 80 MeV.

Moreover, $\gamma$
samples from different processes are selected to measure the time resolution
at various energies; it scales according to the law
$\sigma_{t}$=(54/$\sqrt(E(GeV) \oplus$ 50) ps.

At the present luminosity then KLOE trigger system \cite{trp}
acquires data at a rate of $\sim$ 2.5 khz,
resulting in a throughput to the daq system\cite{dp} of $\sim$ 3
mbytes/s. the data are reconstructed quasi-on line by a dedicated 
computer farm, whose computing power exceeds 5000 specints95.

\section{\KS\ decays}
When a $\phi$ meson decays into two neutral kaons, $c$-invariance
forces the two kaons to be in a \KS-\KL\ state. The observation of
a \KL, therefore, {\it tags} the presence of the \KS\ in the
opposite hemisphere. Moreover, at \DAF\ \KS\ is
produced with a 110 MeV/c momentum and its decay
occurs very close to the interaction point. 

The determination of the ratio of the partial decay widths
into two charged and into two neutral pions is presented. This ratio
is relevant to the main KLOE physical item, the CP violation in the
\KL\ - \KS\ system, since it is a part of the
double ratio from which the CP violation parameter \Repsp\ is derived. 
Moreover it is of
interest for low energy hadron phenomenology, especially if the
radiation of soft photons in the charged decay is properly taken
into account\cite{cdg}. We present also the measurement of the 
branching ratio of the decay \Sep.

These data correspond to a integrated luminosity of $\sim$ 17
pb$^{-1}$, acquired with the detector in near perfect and stable
conditions.

As \KS\ tagging strategy, one can look for a \KL\ interaction in the EmC
 by searching for an EmC cluster of energy
deposit compatible with that due to a slowly moving ($\beta
\approx$ 0.22) neutral particle (called a 'KCRASH' event).
Actually, more than one half of the \KL 's reach the calorimeter
before they decay. Thus, the 'KCRASH' tag  provides a particularly
clean, abundant \KS\ mesons sample.

\subsection{\Kpm\ and \K00\ decays}

The \KS\ decaying into two  neutral pions events are selected by
requiring the presence of at least three EmC clusters with a
timing compatible with the hypothesis of being due to prompt
photons (within 5 $\sigma$'s), and energy larger than 20 MeV. 
The efficiency for detecting a
photon of given energy/angle has to be properly evaluated: this is
done with real data using $\gamma$'s in the decays $\phi
\rightarrow \pi^{+}\pi^{-}\pi^{0}$ as a control sample. 
The final selection efficiency for the \K00\ decay channel is
$\epsilon_{00}$=(90.1$\pm$0.5)$\%$, dominated by acceptance.
the selection of \Kpm\ events proceeds through asking for two
oppositely charged tracks with polar angle  in the interval
30$^{\circ} < \theta <$ 150$^{\circ}$, originating in a cylinder
of 4 cm radius and 10 cm length around the interaction point. A
further cut is applied on the measured momenta to remove the
residual background due to charged kaon tracks: 120 $<$ p(MeV/c)
$<$ 300. 

The track reconstruction efficiency is measured in momentum and polar angle
bins from data subsamples. The final selection efficiency is
$\epsilon_{+-}$=(58.5$\pm$0.1)$\%$, again dominated by acceptance.

The trigger efficiency is determined with real data for both decay
types. it is ( 99.69 $\pm$ 0.03)$\%$ for the neutral decay and
( 96.5 $\pm$ 0.1 )$\%$ for the charged one. The above figure includes also
the probability for having at least one good cluster to determine the
t$_{0}$ of the event.
Background levels are kept well below 1$\%$ for both decay types.
The final result is \cite{ksratio}
\begin{equation}
\frac{\Gamma( k^{0}_{s} \rightarrow \pi^{+}\pi^{-})}
{ \Gamma(  k^{0}_{s} \rightarrow \pi^{0}\pi^{0})} 
= 2.239 \pm 0.003_{\rm stat}\pm 0.015_{\rm syst} 
\end{equation} 
to be compared with the present PDG (particle data group) 
value 2.197 $\pm$ 0.026. It is
noteworthy that our measurement is a fully inclusive
measurement of \Kpmg, whereas the experiments quoted in the PDG do
not have a clear indication of the cut on the photon energy and the
corresponding efficiency.

\subsection{\Sep\ decays}
In order to search for \Sep\ decay candidates, events with a KCRASH and two
oppositely charged tracks from the interaction region are initially selected.
events are then rejected if the two tracks
invariant mass (in the pion hypothesis) and the resulting \KS\ momentum
in the $\phi$ rest frame  are compatible with those expected for a \Kpm\ decay,
which is three orders of magnitude more abundant.
According to monte carlo, this preselection has an efficiency, 
after the tag, of $\sim$ 62.4$\%$ on the signal.

In order to perform a time of flight identification of the charged
particles, both tracks are required to be associated with an EmC
cluster. The acceptance for such a request, estimated by monte
carlo, is (51.1 $\pm$ 0.2) $\%$. The efficiency on the signal,
estimated by means of \KL 's decaying into $\pi$e$\nu$ before the
dc internal wall, is (82.0$\pm$0.7)$\%$ .

All efficiencies releted to trigger, track to cluster association, 
good t$_{0}$ determination, are measured
directly on data, making use of \Len, $\phi \rightarrow
\pi^{+}\pi^{-}\pi^{0}$ and \Kpm subsamples. The product of these
efficiencies is (81.7 $\pm$ 0.5) $\%$.

\begin{figure}%1
\begin{center}
\mbox{\psfig{file=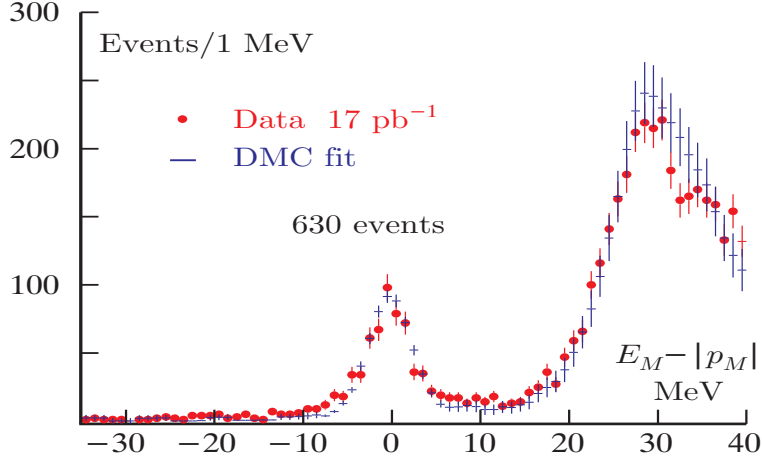,height=6cm,width=10cm}}
\end{center}
\caption{Distribution of the difference between missing energy and missing
momentum for \Sep\ candidates. The peak at zero is the signal.
The distribution is fit to
the monte carlo of signal and background in the range -40 MeV + 40 MeV.}
\label{fitse}
\end{figure}

Once all the particles identities and momenta are known, the event can be
 kinematically closed.
The \KS\ momentum is estimated making use of the measured direction of the \KL\
and of the $\phi$ 4-momentum. The missing energy and momentum of the
\KS\ -$\pi$-e system are then computed. Their difference is distributed as
 in figure \ref{fitse}; it must be equal zero for the signal. Data are
 fitted using mc spectra for both signal and background.

The measured yield is N(\Sep) = 627 $\pm$ 30 events, for a total
efficiency of (21.8 $\pm$ 0.3) $\%$. The total number of events is
then divided by the number of observed \Kpm\ events, to give \cite{kssemi}
BR(\Sep) = (6.92 $\pm 0.34_{stat} \pm 0.15_{syst}$
)$\times$10$^{-4}$. 
This result should be compared with BR(\Sep) = (6.70 $\pm$
0.07)$\times$10$^{-4}$, obtained using the \KS\ and \KL\ lifetimes
and $\Gamma$(\Sep)=$\Gamma$(\Len) which follows from $CPT$ invariance 
and $\Delta S=\Delta Q$. A precise measurement of this decay is a test of the
$\Delta S=\Delta Q$ rule. 

\section{ Radiative $\phi$ decay : 
BR($\phi\rightarrow \eta^{'}\gamma$) }

The branching ratio of the decay  $\phi\rightarrow \eta^{'}\gamma$ is 
particularly interesting  since its value can probe the $|s\bar{s}\rangle$  
and gluonium content of the $\eta^{'}$.
In particular, the ratio of its value to the one of 
$\phi\rightarrow \eta\gamma$ can be related to the 
$\eta - \eta^{'}$ mixing parameters, offering an 
important point of  comparison for the chiral perturbation theory.

The decay chain considered in KLOE for the $\eta^{'}$ selection is 
$\phi\rightarrow \eta^{'}\gamma$ with $\eta^{'} \rightarrow \pi^+\pi^-\eta$
and then $\eta \rightarrow \gamma\gamma$\cite{etaprimo}. 
The $\eta$ is instead selected in the chain
$\phi\rightarrow \eta\gamma$ with $\eta \rightarrow \pi^+\pi^-\pi^0$
and $\pi^0 \rightarrow \gamma\gamma$.

In both cases the final state is $\pi^+\pi^-\gamma\gamma\gamma$, resulting in 
several systematic errors cancellation in the
measurement of the ratio 
BR($\phi\rightarrow \eta^{'}\gamma$)/br($\phi\rightarrow \eta\gamma$). 
For both channels events are selected requiring the presence of three 
prompt photons and two
tracks of opposite charge with a vertex near the interaction region. 
Then a preliminary kinematic fit is performed requiring total energy and 
momentum conservation, the constraint $\beta$ = 1 for all photons, without 
any invariant mass constraint on intermediate particles. 
Simple kinematic cuts eliminate the background mainly due to
$\phi \rightarrow \pi^+\pi^-\pi^0$  
and to $\phi\rightarrow K_L K_S \rightarrow \pi^+\pi^-\pi^0\pi^0$
events.

%the distribution of the invariant mass of $\pi^+\pi^-\gamma\gamma$
% for the $\phi\rightarrow \eta^{'}\gamma$ events is shown in fig. \ref{etaf}, 
%where a small residual background is observed in the $\eta^{'}$ peak region. 
The number of signal events
%,as obtained from the fit, 
is $N_{ev} = 124 \pm 12_{stat} \pm 5_{syst}$. The ratio 
BR($\phi\rightarrow \eta^{'}\gamma$)/BR($\phi\rightarrow \eta\gamma$) 
is obtainedby normalizing to the number of
$\phi\rightarrow \eta\gamma$ observed events and correcting for 
detection efficiency taken from MC. The KLOE preliminary result 
is then \cite{etaprimo}:

\begin{equation}
 \frac{BR \left( \phi\rightarrow \eta^{'}\gamma \right)}
      {BR\left( \phi\rightarrow \eta\gamma \right)} = 
\left( 5.3 \pm 0.5_{stat} \pm 0.4_{syst}\right) \times 10^{-3}
\end{equation}

Making use of BR($\phi \rightarrow \eta \gamma$) value from PDG, 
 a preliminary KLOE value is obtained \cite{etaprimo}: 
\begin{equation}
BR(\phi\rightarrow \eta^{'}\gamma) = 
\left( 6.8 \pm 0.6_{stat} \pm 0.5_{syst}\right) \times 10^{-5}
\end{equation}

\section{ Radiative $\phi$ decay : 
BR($\phi\rightarrow f_0\gamma \rightarrow \pi^0\pi^0\gamma$) 
and BR($\phi\rightarrow a_0\gamma \rightarrow \eta\pi^0\gamma$)}

\begin{figure}%1
\begin{center}
\mbox{\psfig{file=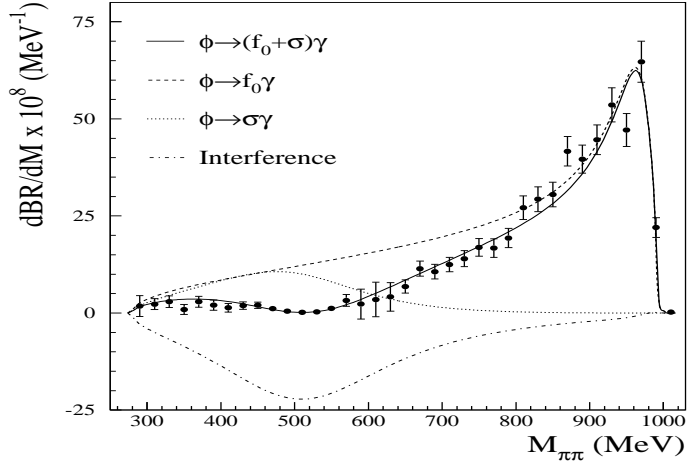,height=7cm,width=10cm}}
\end{center}
\caption{$dBR/dM$ as a function of $M_{\pi\pi}$ from $\phi
\rightarrow f_{0}\gamma$ and $f_{0}\rightarrow \pi^0\pi^0$. Points are
data. The solid line is the fit resulting from assuming a $f_{0}$,
a $\sigma$ and the interference between them. contributions of
individual terms are also shown.} 
\label{f0mass}
\end{figure}

The several models proposed to explain the nature of the
$f_0$ and $a_0$ mesons (ordinary qq meson, qqqq state, KK molecule)
make different predictions\cite{f0,a0} of BR($\phi\rightarrow f_0\gamma$), 
BR($\phi\rightarrow a_0\gamma$) and their ratio. 
Therefore a precise measurement of
BR($\phi\rightarrow f_0\gamma$)/BR($\phi\rightarrow a_0\gamma$) ratio 
would clarify  the interpretation on the nature of these mesons. 
The $f_0$ meson at KLOE has been studied in the following decay chains :
$\phi\rightarrow f_0\gamma$  with $f_0\rightarrow\pi^0\pi^0$. The $a_0$ has 
been investigated in the channel: $\phi\rightarrow a_0\gamma$ 
with the  $a_0 \rightarrow \eta \pi^0$ and the following decays of the $\eta$:
$\eta \rightarrow \gamma\gamma$ and $\eta \rightarrow \pi^+\pi^-\pi^0$.
all these final states have 5 $\gamma$'s. 

With the exception of the very clean case of the $a_0$ production where 
the final state is $\pi^+\pi^- 5 gamma$, the main backgrounds 
come from the resonant process $\phi \rightarrow \rho^0\pi^0$ 
with $\rho^0 \rightarrow \pi^0\gamma$ 
or $\rho^0 \rightarrow\eta\gamma$, and the continuum process 
$e^+e^- \rightarrow \omega\pi^0$ with $\omega \rightarrow \pi^0\gamma$ 
or $\omega \rightarrow \eta\gamma$. 

The events are identified by requiring five prompt photons 
within loose acceptance cuts. 
Then kinematic fits are performed requiring total energy and 
momentum conservation with the additional constraint 
$\beta=1$ for all photons. 
Background events are isolated by means of cuts on kinematic
variables and on the $\chi^2$ probability values 
obtained in the kinematic fits.

The overall detection efficiency is obtained by 
tuning the MC to reproduce the observed
$M_{\pi^0\pi^0}$ and $M_{\eta\pi^0}$
invariant mass distributions in case of 
$\phi\rightarrow\pi^0\pi^0\gamma$ and $\phi \rightarrow \eta\pi^0\gamma$
events, respectively. 

In the $\phi \rightarrow \pi^0\pi^0\gamma$ mode the $M_{\pi^0\pi^0}$ 
spectrum can have contribution by the process $\phi \rightarrow S\gamma$,
 with the scalar meson $S \rightarrow \pi^0\pi^0$, and by 
$\phi \rightarrow \rho^0\pi^0$, with $\rho^0 \rightarrow \pi^0\gamma$\cite{f0}.
We considered the contribution of the $f_0$ and $\sigma$\cite{sigma} scalar
 mesons to fit the dipion spectrum, as shown in fig \ref{f0mass}.
Is clearly visible the negative interference with the $\sigma$ term, while
is negligible the contribution from the $\phi \rightarrow \rho^0\pi^0$ channel.
The following results are obtained integrating the $f_0$ component of
the spectrum\cite{f0}: 
\begin{equation}
BR\left( \phi \rightarrow f_0\gamma \rightarrow \pi^0\pi^0\gamma \right) 
= \left(1.49 \pm 0.07\right) \times 10^{-4} 
\end{equation}

\begin{figure}%1
\begin{center}
\mbox{\psfig{file=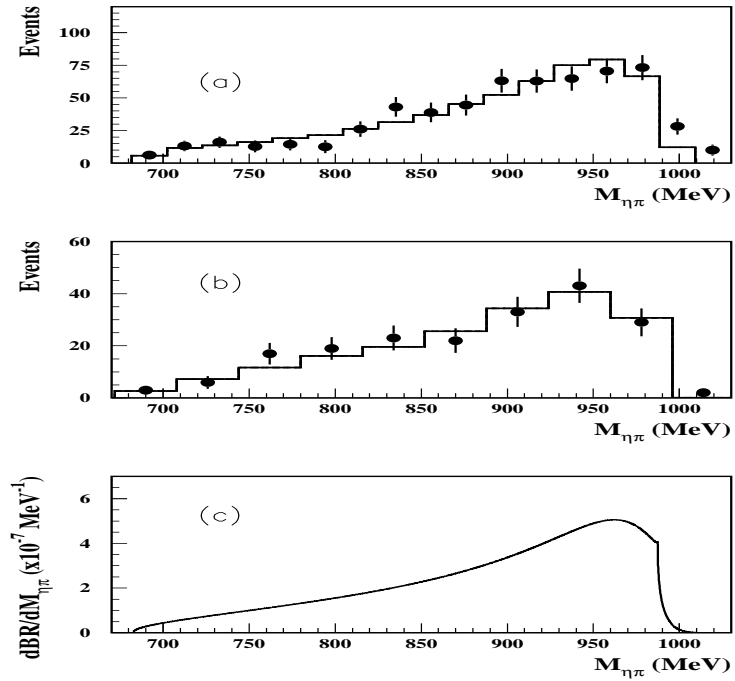,height=10cm,width=11cm}}
\end{center}
\caption{The
background subtracted $M_{\eta\pi}$ spectra resulting from $\phi
\rightarrow a_{0}\gamma$ and $a_{0}\rightarrow \eta\pi$. Top plot
points are data, for the case where $\eta\rightarrow 2\gamma$,
middle plot for case where 
$\eta \rightarrow \pi^+\pi^-\pi^0$. Histograms are results of the
fit and the theoretical curve for the $a_0$.} 
\label{a0mass}
\end{figure}

In the $\phi \rightarrow \eta\pi^0\gamma$ case the two $m_{\eta\pi^0}$
spectra obtained for the samples where $\eta \rightarrow \gamma\gamma$ and 
$\eta \rightarrow \pi^+\pi^-\pi^0$ (fig \ref{a0mass}) are fit together
setting the $a_0$ mass parameter = 984.8 MeV from PDG.
From this fit we obtain \cite{a0}: 
\begin{equation}
BR\left( \phi \rightarrow a_0\gamma, a_0 \rightarrow \eta\pi^0 \right) = 
\left( 7.4 \pm 0.7 \right) \times 10^{-5} 
\end{equation}
The presented results on the $f_0$ and $a_0$ scalars can be put together
to give a ratio\cite{a0}:
\begin{equation}
\frac{BR(\phi\rightarrow f_0\gamma)}{ BR(\phi\rightarrow a_0\gamma)}
 = 6.1 \pm 0.6
\end{equation}

\section{Conclusions}
The KLOE detector at \DAF\ has collected about 200 pb$^{-1}$ of
data so far. In this report studies on $\phi$
radiative decays have been presented, together with precision
measurements on two \KS\ decay channels, based on the statistics
accumulated in year 2000. Charged kaon decays as well as other
\KS\ decays, such as \KS\ $\rightarrow 3\pi^{0}$ or \KS\
$\rightarrow \gamma\gamma$ are under investigation. Results on
these last topics are expected in a short time.

%\section*{Acknowledgments}
%We thank the DA$\Phi$NE team for their efforts in maintaining low
%background running conditions and their collaboration during all
%data-taking. This work was
%supported in part by the German Federal Ministry of Education and Research 
%(BMBF) contract 06-KA-957; 
%by Graduiertenkolleg 'H.E. Phys.and Part. Astrophys.' of 
%Deutsche Forschungsgemeinschaft, Contract No. GK 742;
%by INTAS, contracts 96-624, 99-37 and by TARI, contract HPRI-CT-1999-00088.

\end{document}